\documentclass[preprint,12pt,showpacs,aps]{revtex4}
\usepackage{graphicx}
\usepackage{amssymb}
\begin{document}

\title{Quantum effect on luminosity-redshift relation}
\author{Li-Fang Li}
  \affiliation{Department of Physics, Beijing Normal University, Beijing 100875, China}

\author{Jian-Yang Zhu}
\thanks{Author to whom correspondence should be addressed}
  \email{zhujy@bnu.edu.cn}
  \affiliation{Department of Physics, Beijing Normal University, Beijing 100875, China}

\begin{abstract}
There are many different proposals for a theory of quantum gravity.
Even leaving aside the fundamental difference among theories such as
the string theory and the non-perturbative quantum gravity, we are
still left with many ambiguities (and/or parameters to be
determined) with regard to the choice of variables, the choice of
related groups, etc. Loop quantum gravity is also in such a state.
It is interesting to search for experimental observables to
distinguish these quantum schemes. This paper investigates the loop
quantum gravity effect on luminosity-redshift relation. The quantum
bounce behavior of loop quantum cosmology is found to result in
multivalued correspondence in luminosity-redshift relation. And the
detail multivalued behavior can tell the difference of different
quantum parameters. The inverse volume quantum correction does not
result in bounce behavior in this model, but affects
luminosity-redshift relation also significantly.
\end{abstract}


\maketitle

\section{Introduction}
In recent years, we witnessed rapid development in quantum gravity
theories, especially in the string theory and the loop quantum
gravity theory. Both of them have produced many important results.
For loop quantum gravity, area and volume operators have been
quantized \cite{rovelli95,ashtekar97,ashtekar98a,thiemann98}. The
entropy of black holes \cite{rovelli96} can be calculated from
statistical mechanics. In addition, as a successful application of
loop quantum gravity to cosmology, loop quantum cosmology (LQC) has
an outstanding result---replacing the big bang spacetime singularity
of cosmology with a big bounce \cite{bojowald01}. LQC also gives a
quantum suppression of classical chaotic behavior near singularities
in the Bianchi-IX models \cite{bojowald04a,bojowald04b}.
Furthermore, it has been shown that non-perturbative modification of
the matter Hamiltonian leads to a generic phase of inflation
\cite{bojowald02,date05,xiong1}. We also know that there are many
alternative ways for quantum gravity, including the string theory
and non-perturbative quantum gravity. Within non-perturbative
quantum gravity only, we have also many choices for quantization,
with freedom in, for example, the choice of variables and the choice
of the related groups. Thus any contribution from experimental
observation would be all the more valuable.

In cosmology, the data accumulated are plenty, actually more than
what theory can explain. Among the observations, the
luminosity-redshift relation of type Ia supernovae (SNIa) suggests
that the universe has entered a phase of accelerating expansion and
that the universe is spatially flat \cite{perlmutter99}. It is thus
interesting to see if we can use this observable to verify any
quantum gravity schemes, or use a gedanken experiment to test the
different behaviors of different schemes. This is the topic of this
paper. Based on the effective LQC theory, we look into the effect of
quantum correction on the luminosity-redshift relation, and test the
different choices involved in LQC, which have been causing
ambiguity. Classically, scalar field is used to explain dark
matter/energy, and the scalar field model fits the experiment data
well \cite{matos00}. Therefore we use this model as the classical
reference for investigating the quantum effects.

This paper is organized as follows. The next section, we give a
brief review of effective LQC theory. In this paper we consider only
the holonomy correction and the inverse volume correction. In
Sec.\ref{Sec.3}, we use a massless scalar field to model the dark
matter/energy, and derive the classical luminosity-redshift relation
of this model. For comparison with LQC, we adopt the formalism of
effective LQC to express this classical dynamics. In
Secs.\ref{Sec.4} and \ref{Sec.5}, we study the quantum effects,
paying much attention to the $l$-parameter in the $\bar{\mu}$
scheme. We conclude the paper in the last section. Through out this
paper we adopt $c=G=\hbar=1$ which results in the Hubble constant
$H_0\approx 1.2\times10^{-61}$.

\section{framework of effective LQC}\label{Sec.2}
Based on the assumptions of cosmological principle and that the
universe is spatially flat, the metric of the related spacetime is
described by FRW metric
\begin{eqnarray}
ds^2=-dt^2+a^2(dx^2+dy^2+dz^2),
\end{eqnarray}
where $a$ is the scale factor of the universe, which only depends on
$t$ due to homogeneity of our universe. The classical Hamiltonian
for the system we considered in this paper is given by
\begin{equation}
H_{cl}=-\frac 3{8\pi \gamma ^2}\sqrt{p}c^2+H_M\left( p,\phi \right).
\end{equation}
Here we have adopted the Ashtekar variables in loop quantum gravity.
The phase space is spanned by the generalized coordinates $c=\gamma
\dot{a}$ and the generalized momentum $p=a^2$. $\gamma=0.2375$ is
the Barbero-Immirzi parameter. $H_M$ denotes the Hamiltonian of the
matter part and $\phi$ denotes the matter field. Together with the
Poisson bracket for the gravity part, which is defined for any two
functions $f$ and $g$ on phase space as
\begin{eqnarray}
\{f,g\}:=\frac{8\pi\gamma}{3}\left(\frac{\partial f}{\partial
c}\frac{\partial g}{\partial p}-\frac{\partial f}{\partial
p}\frac{\partial g}{\partial c}\right),
\end{eqnarray}
we can get the corresponding canonical equations.

Correspondingly, the effective Hamiltonian in LQC is given by
\cite{ashtekar06}
\begin{equation}
H_{eff}=-\frac 3{8\pi \gamma ^2\bar{\mu}^2}\sqrt{p}\sin ^2\left(
\bar{\mu} c\right) +H_M\left( p,\phi \right). \label{hamilton}
\end{equation}
If we consider the inverse volume quantum correction, $H_M$ will
change correspondingly. We will describe more in Sec.\ref{Sec.5}.
The variable $\bar{\mu}$ corresponds to the dimensionless length of
the edge of the elementary loop and is given by
\begin{equation}
\bar{\mu}=\xi p^{-l},
\end{equation}
where $\xi>0$ is a constant and depends on the particular scheme in
the holonomy corrections. According to the idea of area quantization
and the requirement of area gap, we have \cite{hrycyna09}
\begin{equation}
\xi^2=2\sqrt{3}\pi\gamma l_p^2,
\end{equation}
where $l_p=1$ is the Planck length. $l$ is an ambiguous parameter in
LQC. Considerations of the lattice states place an restriction that
$l\in(0,0.5]$ \cite{resl1}. The anomaly cancellation and the
positivity of the graviton's effective mass requires
$l\in[0.1319,2.5]$ \cite{resl2}. But we are still not able to fix
this ambiguity parameter theoretically. Later in this paper we will
test the influence of different values of $l$ on the
luminosity-redshift relation.
\section{Classical scalar field model} \label{Sec.3}
In this section we will investigate the classical dynamics of the
universe using scalar field to model the dark matter/energy. The
universe is nearly homogeneous and isotropic, with roughly $4\%$ of
the matter content being ordinary matter, $22\%$ being cold dark
matter and $74\%$ being dark energy. Until now, we do not know the
nature of the cold dark matter and the dark energy, here we use a
massless scalar field to model the cold dark matter and  the dark
energy, and ignore the ordinary matter part. The Hamiltonian of the
classical scalar field model is
\begin{eqnarray}
H_{cl}&=&-\frac{3}{8\pi
\gamma^2}c^2\sqrt{p}+\frac{p_{\phi}^2}{2p^{3/2}}+p^{\frac{3}{2}}V(\phi),
\end{eqnarray}
where $p_{\phi}$ is the conjugate momentum of the free massless
scalar field $\phi$. Following \cite{matos00} we take
\begin{eqnarray}
V=V_0e^{-\sqrt{\frac{16\pi}{\lambda}}\phi},
\end{eqnarray}
where $V_0$ is some constant. The equations of motion in this
classical case are given by Hamilton's equations:
\begin{eqnarray}
\dot{c}&=&-\frac{1}{2\gamma\sqrt{p}}c^2-2\pi\gamma\frac{p^2_\phi}{p^{5/2}}+
4\pi\gamma V_0p^{\frac{1}{2}}e^{-\sqrt{\frac{16\pi}{\lambda}}\phi},\nonumber\\
\dot{p}&=&\frac{2}{\gamma}c\sqrt{p},\nonumber\\
\dot{\phi}&=&p^{-3/2}p_{\phi},\nonumber\\
\dot{p_{\phi}}&=&\sqrt{\frac{16\pi}{\lambda}}V_0p^{\frac{3}{2}}e^{-\sqrt{\frac{16\pi}{\lambda}}\phi}.\nonumber
\end{eqnarray}

The Friedmann equation reads
\begin{eqnarray}
H^2=\frac{8\pi}{3}\rho,
\end{eqnarray}
where $\rho=\frac{\dot{\phi}^2}{2}+V(\phi)$. The exact solution to
the model described above is
\begin{eqnarray}
a&=&\left(\frac{t}{\bar{t}}\right)^\lambda,\label{exact}\\
\phi&=&\sqrt{\frac{1}{4\pi\lambda}}\ln a,
\end{eqnarray}
with $\bar{t}=\sqrt{\frac{\lambda(3\lambda-1)}{8\pi V_0}}$. And the
parameter of the state equation is $\omega=\frac{2}{3\lambda}-1$. We
will see in the following that $\omega$ is a negative constant, to
be consistent with the property of cold dark matter and dark energy,
for which the equation of state parameters are 0 and -1
respectively.

\begin{figure}[ht]
\begin{tabular}{c}
\includegraphics[width=0.5\textwidth]{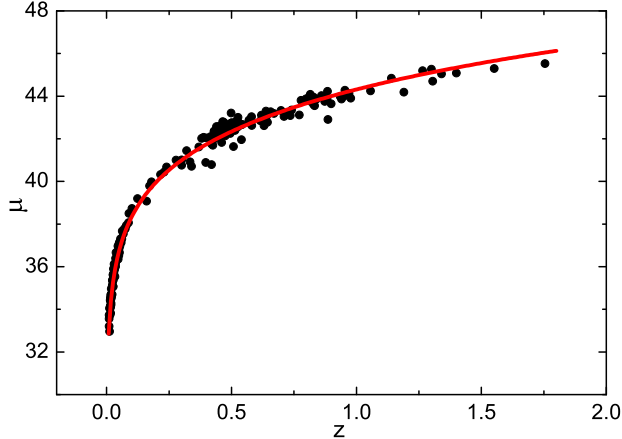}
\end{tabular}
\caption{Fit of the theoretical luminosity-redshift relation (line),
as described in Eq.(\ref{eq15}), to the experimental data for type-I
supernova reported in \cite{reiss04} (solid circles).}\label{fig1}
\end{figure}

The luminosity distance is a way of expressing the amount of light
received from a distant object. When we receive a certain flux from
an object, we can calculate the luminosity distance between us and
the object, assuming the inverse square law for the reduction of
light intensity with distance holds. Because of the expansion of the
universe, the number of photons in unit volume of a sphere shell
will decrease $\propto \frac{a_0}{a}=(1+z)$, where $z$ is the
redshift factor. Considering the cosmological redshift effect, the
individual photons will lose energy $\propto (1+z)$. Based on the
above inverse square law assumption, we get luminosity distance
\begin{eqnarray}
d_L=a_0(1+z)\int_{0}^{t}\frac{dt}{a(t)}. \label{eq12}
\end{eqnarray}
Given the above exact solution, we have
\begin{eqnarray}
d_L=\frac{(1+z)\lambda}{H_0(1-\lambda)}\left[1-(1+z)^{1-\frac{1}{\lambda}}\right].
\label{eq13}
\end{eqnarray}
According to the astronomical convention, we adopt the logarithmic
measure of the luminosity instead of luminosity itself in presenting
our result,
\begin{eqnarray}
\mu=5\log_{10}d_L-286.4, \label{eq14}
\end{eqnarray}
here -286.4 comes from our units $c=G=\hbar=1$ \footnote{In usual
literature, ones use Mpc as unit of $d_L$ which results in 25.
Useing the Planck units gives the result of -286.4, since
$l_p\approx
%1.6\times 10^{-35}$m$\approx
5.24\times 10^{-58}$Mpc.
%1Mpc$ = 3.08\times 10^{22}$m, $H_0\approx2.5\times10^{-18}$s${}^{-1}=1.35\times10^{-61}t_p^{-1}$.
%$t_p\approx5.4\times10^{-44}$s.
}. Using Eq.(\ref{eq13}) in the above equation, we get
\begin{eqnarray}
\mu=43.17+5\log_{10}\left\{{\frac{\lambda(1+\lambda)}{1-\lambda}\left[1-(1+z)^{1-\frac{1}{\lambda}}\right]}\right\}.
\label{eq15}
\end{eqnarray}
Fitting $\lambda$ in this equation to the experimental data reported
in \cite{reiss04}, we get $\lambda=2.38$ which results in the
parameter of the state equation $\omega=-0.72$. For cold dark matter
and dark energy the total pressure can be expressed as
\begin{eqnarray}
P=P_{M}+P_{E}=-\rho_{E}=-0.72\rho,
\end{eqnarray}
where subfix $M$ means the cold dark matter, and $E$ means dark
energy. In the above equation, we have used $\omega_{M}=0$ and
$\omega_{E}=-1$. Thus our fitted result $\omega=-0.72$ is consistent
with roughly $74\%$ of our universe being dark energy. As
Fig.\ref{fig1} shows, the resulted line lies closely with the
experimental data.

In order to investigate the quantum effects, we turn to the quantum
region of the universe. We introduce a parameter
$\alpha:=\frac{\rho_{M}}{\rho_c}$, the ratio between the matter
density and the quantum critical density
($\rho_c=\frac{\sqrt3}{16\pi^2\gamma^3}\approx0.82$), to indicate
how close we are to the quantum region. Given $\alpha$, we can
determine $\rho_M$ and then the time $t$ which gives all the
dynamical variables according to the above exact solution
(\ref{exact}). In the rest of this paper, we regard $t$ as the
``present time", to investigate the quantum effects on
luminosity-redshift relation. Since the constant of  $286.4$ is
irrelevant to us for comparing with the classical
luminosity-redshift relation, we will ignore it in the following,
and will define the luminosity-redshift relation simply as
\begin{eqnarray}
\mu=5\log_{10}d_L.
\end{eqnarray}

\section{holonomy quantum correction of the scalar field model}
\label{Sec.4} Based on the effective description of quantum dynamics
for LQC as described in Sec.\ref{Sec.2}, the Hamiltonian can be
written as \cite{ashtekar06,bentivegna08}
\begin{eqnarray}
H_{eff}^{hc}&=&-\frac{3}{8\pi\gamma^2
\bar{\mu}^2}p^{1/2}\sin^2(\bar{\mu}c)+\frac{p_{\phi}^2}{2p^{3/2}}+p^{\frac{3}{2}}V_0
e^{-\sqrt{\frac{16\pi}{\lambda}}\phi},
\end{eqnarray}
for the holonomy quantum correction of the scalar field model
considered in this work. In the limit of $\bar{\mu}\rightarrow 0$,
this Hamiltonian reduces to the standard classical one. The
dynamical equations are given by
\begin{eqnarray}
\dot{c}&=&-\frac{1}{2\gamma\bar{\mu}^2\sqrt{p}}\sin^2(\bar{\mu}c)+2l\frac{1}{\gamma\bar{\mu}\sqrt{p}}\sin(\bar{\mu}c)\cos(\bar{\mu}c)c\nonumber\\
&-&2l\frac{1}{\gamma\bar{\mu}^2\sqrt{p}}\sin^2(\bar{\mu}c)-2\pi\gamma
p_{\phi}^2 p^{-\frac{5}{2}}+
4\pi\gamma V_0p^{\frac{1}{2}}e^{-\sqrt{\frac{16\pi}{\lambda}}\phi},\\
\dot{p}&=&\frac{2}{\gamma}\frac{\sqrt{p}}{\bar{\mu}}\sin(\bar{\mu}c)\cos(\bar{\mu}c),\\
\dot{\phi}&=&p^{-\frac{3}{2}}p_{\phi},\\
\dot{p}_{\phi}&=&\sqrt{\frac{16\pi}{\lambda}}V_0p^{\frac{3}{2}}e^{-\sqrt{\frac{16\pi}{\lambda}}\phi}.
\end{eqnarray}
\begin{figure*}[th]
\begin{tabular}{cc}
\includegraphics[width=0.5\textwidth]{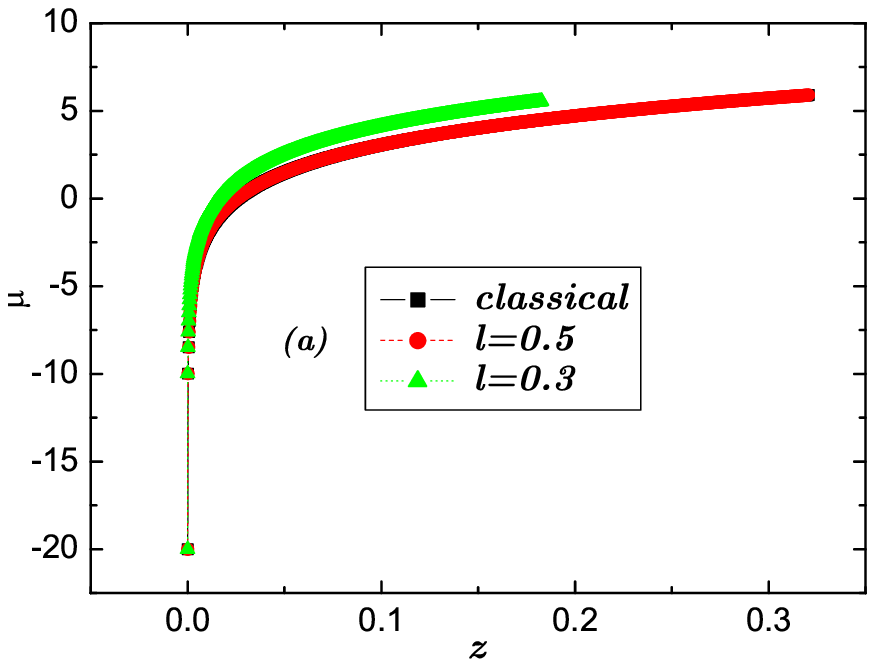}&
\includegraphics[width=0.5\textwidth]{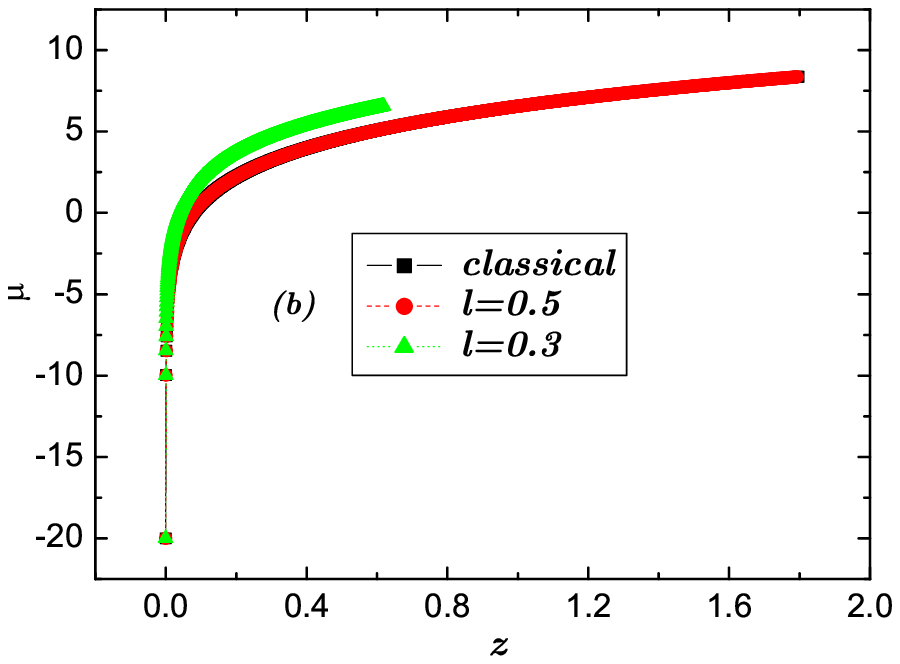}\\
\includegraphics[width=0.5\textwidth]{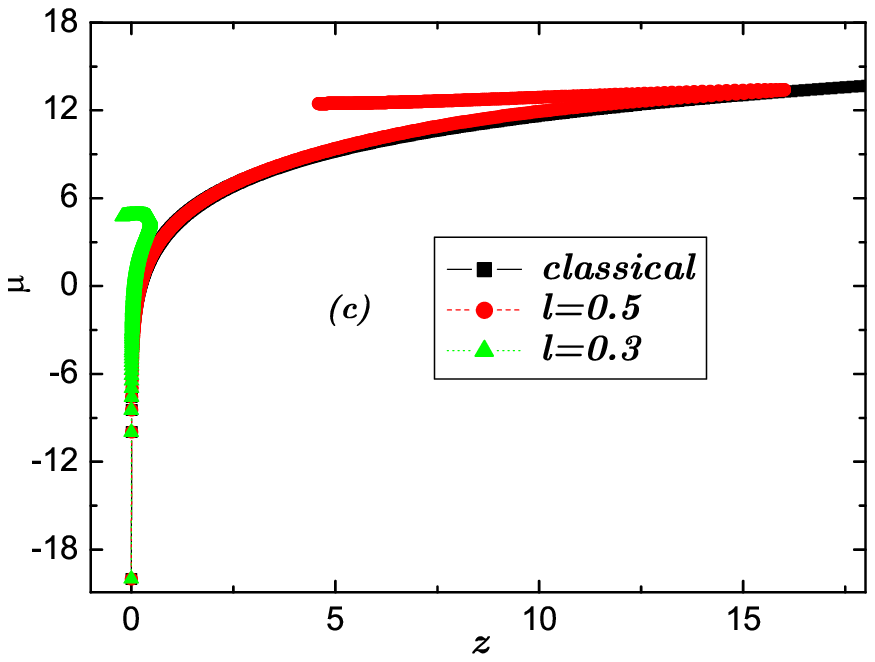}&
\includegraphics[width=0.5\textwidth]{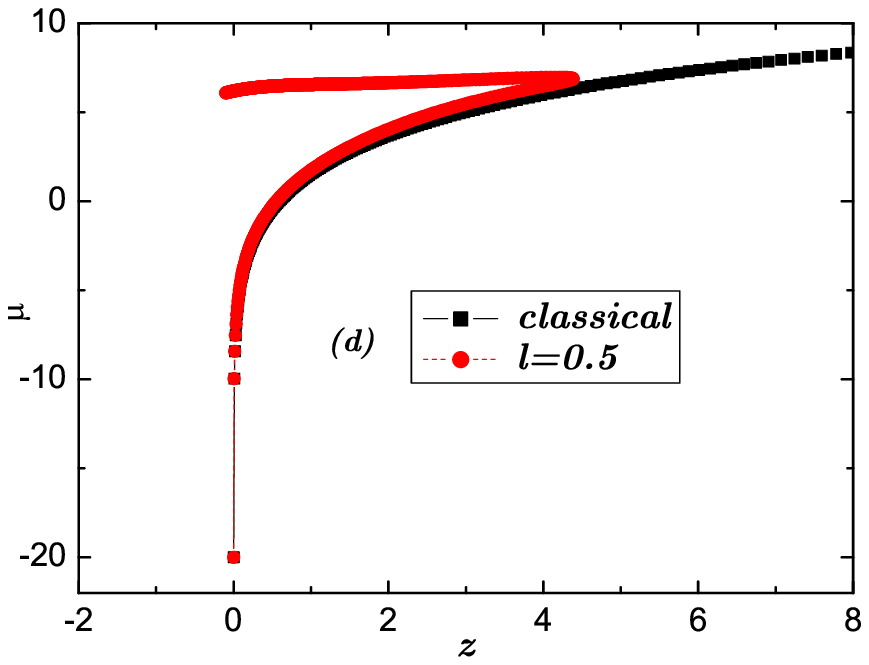}
\end{tabular} \caption{Comparison of the luminosity-redshift
relation in the classical scenario and in the holonomy quantum
correction scenario. The corresponding parameters $\alpha$ are
0.0001 in (a), 0.001 in (b), 0.01 in (c), and 0.1 in (d). The
holonomy quantum correction gives a multi-valued luminosity-redshift
relation.}\label{fig2}
\end{figure*}
Following the above classical considerations, we adopt the initial
condition $p_\phi(0)$, $p(0)$ and $\phi(0)=0$ from the classical
scenario. Then $H_{hc}=0$ is used to determine $c(0)$.
Fig.\ref{fig2} shows the resulted luminosity-redshift relations for
(a) $\alpha=0.0001$, (b) $0.001$, (c) $0.01$, and (d) $0.1$. We can
see that quantum corrections with different parameters $l$ gives
different luminosity-redshift behaviors. As is clear in
Fig.\ref{fig2} (a-c), for larger $l$, it takes a larger $\alpha$ for
the quantum effects to be visible. In (d), the quantum corrected
universe with $l=0.3$ does not admit $\rho_M\leq0.1\rho_c$, so it is
left out from this panel. The most interesting result is that the
quantum correction leads to a multi-valued luminosity-redshift
relation. This multi-valued behavior is actually a result of the
quantum bounce when the universe becomes very small. So this
multi-valued behavior is a common result for all quantum bounce
cosmology models. We expect that the quantum effects can be
distinguished through luminosity-redshift relation if we can observe
objects far enough. In addition, different values of quantum
parameter  $l$ give different quantum bounce behaviors: for smaller
$l$, the luminosity-redshift Relation turns around at smaller $z$.
Thus we conjecture that the luminosity-redshift relation can be used
to fix $l$.

\section{The inverse volume correction of the scalar field
model}\label{Sec.5}

\begin{figure*}[ht]
\begin{tabular}{cc}
\includegraphics[width=0.5\textwidth]{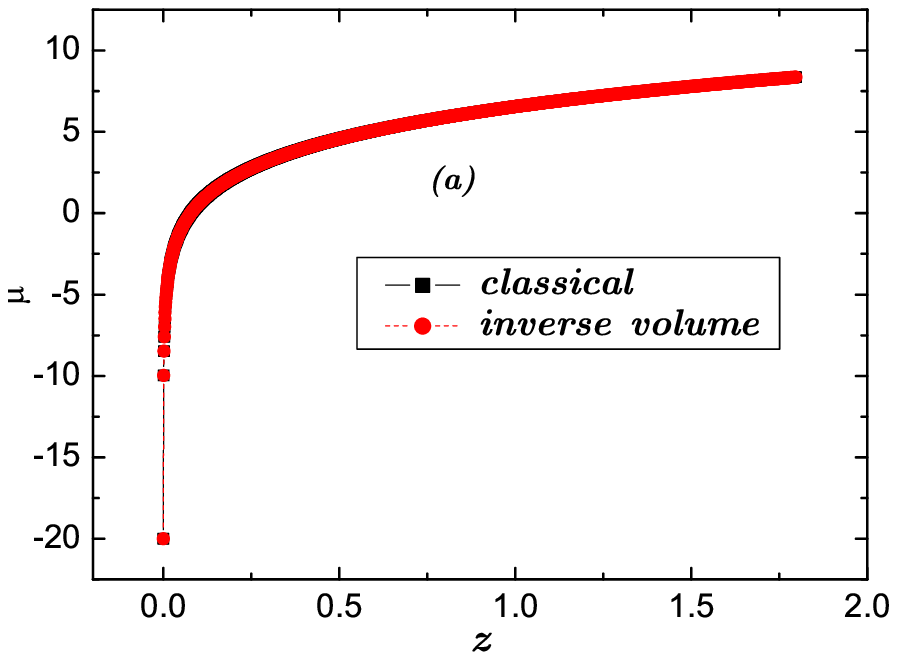}&
\includegraphics[width=0.5\textwidth]{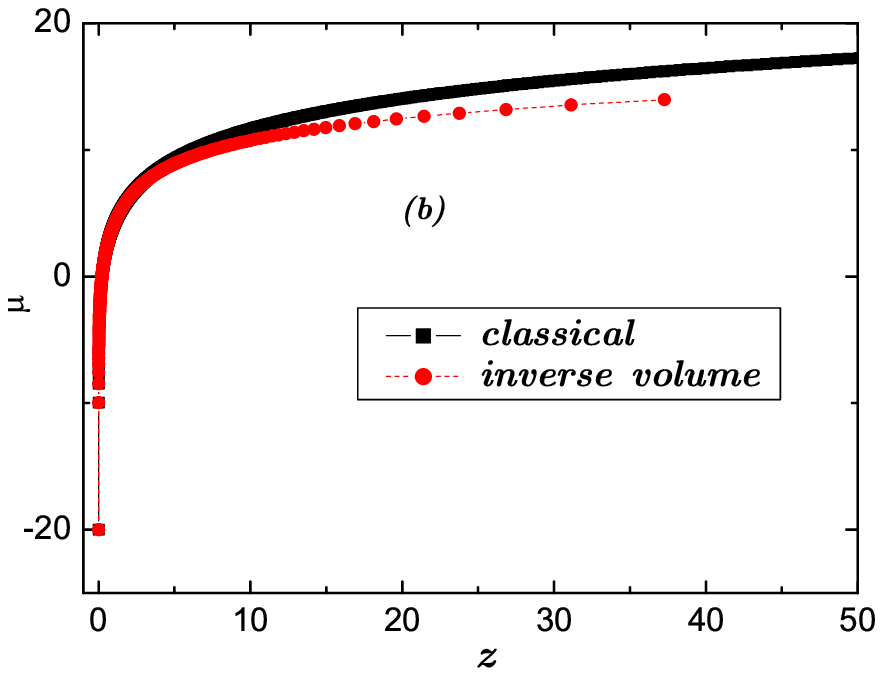}
\end{tabular} \caption{Comparison of the luminosity-redshift
relation in classical scenario and the inverse volume quantum
correction scenario, for $\alpha = 0.001$ in (a) and $0.01$ in
(b). Compared with the holonomy quantum correction, the inverse
volume quantum correction takes place at smaller universe in our
model.}\label{fig3}
\end{figure*}

If we consider the inverse volume correction in LQC, the effective
Hamiltonian can be written as
\cite{bojowald02a,bojowald02b,tsujikawa04}
\begin{eqnarray}
H_{eff}^{iv}=-\frac{3}{8\pi\gamma^2}c^2\sqrt{p}+\frac{1}{2}\frac{D(q)}{p^{3/2}}p_{\phi}^2+p^{\frac{3}{2}}V_0
e^{-\sqrt{\frac{16\pi}{\lambda}}\phi}.\label{inverse_correction}
\end{eqnarray}
Here $D(q)=(\frac{8}{77})^6
q^{3/2}\{7[(q+1)^{11/4}-|q-1|^{11/4}]-11q[(q+1)^{7/4}-\text{sgn}(q-1)|q-1|^{7/4}]\}^6,
q=p/p_*$, and $p_{*}$ is the scale parameter for the inverse volume
correction to take place. With Eq.(\ref{inverse_correction}), we
investigate the effect of the inverse volume correction on the
luminosity-redshift relation, and compare it with the effect of the
holonomy correction. In the classical cosmology, the scale factor
$a=\sqrt p$ has no direct physical meaning. We can always rescale it
to set the value of $a$ at present time to 1. But in loop quantum
cosmology, especially when we consider the inverse volume
correction, the scale factor has important consequences. When its
value is roughly the Planck length $p_*\approx l_p^2$, the inverse
volume correction will take place. The corresponding dynamical
equations are
\begin{eqnarray}
\dot{c}&=&-\frac{1}{2\gamma\sqrt{p}}c^2+\frac{4\pi\gamma}{3p_*}\frac{d D(q)}{d q}p_{\phi}^2p^{-\frac{3}{2}}\nonumber\\
&-&2\pi\gamma D(q)p_{\phi}^2p^{-\frac{5}{2}}+4\pi\gamma V_0p^{\frac{1}{2}}e^{-\sqrt{\frac{16\pi}{\lambda}}\phi},\nonumber\\
\dot{p}&=&\frac{2}{\gamma}c\sqrt{p},\nonumber\\
\dot{\phi}&=&D(q)p^{-3/2}p_{\phi},\nonumber\\
\dot{p_{\phi}}&=&\sqrt{\frac{16\pi}{\lambda}}V_0p^{\frac{3}{2}}e^{-\sqrt{\frac{16\pi}{\lambda}}\phi},
\end{eqnarray}
where
$\frac{dD(q)}{dq}=(\frac{8}{77})^6\frac{3}{2}\sqrt{q}\{7[(q+1)^{11/4}-|q-1|^{11/4}]-11q[(q+1)^{7/4}-\text{sgn}(q-1)|q-1|^{7/4}]\}^6+
(\frac{8}{77})^6q^{3/2}6
\{7[(q+1)^{11/4}-|q-1|^{11/4}]-11q[(q+1)^{7/4}-\text{sgn}(q-1)|q-1|^{7/4}]\}^5\{\frac{33}{4}[(q+1)^{7/4}-\text{sgn}(q-1)|q-1|^{7/4}]
-\frac{77}{4}q[(q+1)^{3/4}-|q-1|^{3/4}]\}$. Combining the
Hamiltonian constraint with the above dynamical equations, we can
get the Friedman equation as
\begin{eqnarray}
\left(\frac{\dot
a}{a}\right)^2=\frac{8\pi}{3}(\frac{\dot\phi^2}{2D}+V).
\end{eqnarray}
Due to the exponential form of our scalar potential $V$, $\dot a$
will never vanish, which means we can not get quantum bounce with
this inverse volume correction. On the other hand, we will see the
luminosity-redshift relation can also distinguish this quantum
correction from classical behavior. We take the same procedure as in
the case of holonomy correction to test this quantum effect on
luminosity-redshift relation. Compared with the classical scenario,
the resulted luminosity redshift relations are plotted in
Fig.\ref{fig3}. Different from the holonomy quantum correction, the
quantum effects take place in smaller universe, roughly
$\rho_M=0.01\rho_c$. Although this may depend on model, the
interesting point is the luminosity-redshift relation as an
observable can distinguish the quantum effect from classical
behavior.

\section{comments and conclusions}

There are currently different proposals for a theory of quantum
gravity, including the string theory and the loop quantum gravity
theory, among others. They all pass the check of self-consistency.
Even in the loop quantum gravity theory alone, there are many
ambiguities that we do not know the way to eliminate. To a large
extent, this situation results from the lack of association with
experimental data. On the other hand, there are plenty of experiment
data in cosmology awaiting satisfactory explanation. For  example,
we have yet to have a consistent theory to explain the accelerating
expansion of the  universe at the present age.

In this paper we use the luminosity-redshift relation to test LQG
effects. We find this relation to be capable of revealing the
quantum effects and fixing the quantum parameters. LQC predicts a
multi-valued luminosity-redshift relation, as a result of the
quantum bounce behavior. The exact shape of the relation can be used
to fix quantum parameters in LQC. We can also distinguish the
inverse volume quantum correction from the classical picture.
Although some of the predictions made in this paper can not yet be
tested using the experiment data available, it is a good gedanken
experiment at least, to study the ambiguities involved in LQC and
how they can be eliminated. Certainly, this paper touches merely a
small part of the problem with ambiguity in LQC, yet our results
show that the quantum bounce behavior is possibly observable through
the multi-valued shape of luminosity-redshift relation.

\acknowledgements

L.-F. Li thanks Prof. Hwei-Jang Yo to bring her into this
interesting subject. The work was supported by the National Natural
Science Foundation of of China (No.10875012) and the Scientific
Research Foundation of Beijing Normal University.

\end{document}